# Measuring and Discovering Correlations in Large Data Sets

Lijue Liu, Ming Li, Sha Wen
College of Information Science and Engineering
Central South University
Chang Sha, PRC
ljliu@csu.edu.cn, limingcsu@foxmail.com, wensha0103@163.com

*Abstract*—In this paper, a class of statistics named ART (the alternant recursive topology statistics) is proposed to measure the properties of correlation between two variables. A wide range of bi-variable correlations both linear and nonlinear can be evaluated by ART efficiently and equitably even if nothing is known about the specific types of those relationships. ART compensates the disadvantages of Reshef's model in which no polynomial time precise algorithm exists and the "local random" phenomenon can not be identified. As a class of nonparametric exploration statistics, ART is applied for analyzing a dataset of 10 American classical indexes, as a result, lots of bi-variable correlations are discovered.

## I. INTRODUCTION

The unknown laws of nature and society are always hidden among massive data in the form of correlation [1-3], such as the relationship between air quality and the developing level of industry, the associations between economic growth and various factors, and so on [4-6]. A medium-sized database may contain hundreds of variables and tens of thousands of hidden correlations. The efficiency of discoverring the desired correlations depends on the method of correlation assessment. The most commonly used method is the ancient correlation coefficient Pearson's r [7], but it captures only linear relationships and its usefulness is greatly reduced when relationships are nonlinear [8].

In the context of information theory, mutual information (MI) can treat linear and nonlinear relationships relatively fairly, and it seems like to be the most promising solution after Pearson's r [9-11]. But MI's value range is uncertain, and no standard exists to compare MI of samples containing relationships of different types, what's more, estimating samples' MI is very difficult, therefore, MI can not be used to measure correlation directly [12-14].

In 2011, the maximal information coefficient (MIC), which belongs to a larger class of statistics for identifying and classifying relationships, was proposed by Reshef, and it was applied for analyzing data sets in major-league baseball, global health, and lots of relationships were found, causing great repercussions [15]. However, MIC can not be worked out in polynomial time accurately, and Reshef provides only an heuristic approximation algorithm; Besides, when "local random" phenomenon exists, MIC can not effectively distinguishes it; Moreover, MIC behaves not well when sample size is small, especially its anti-noise ability is not strong [16].

Considerring the computational deficiency of MIC, this paper proposes a statistic named ARTMIC (the alternant recursive topology maximum information coefficient, which belongs to a larger class of statistics called ART). ARTMIC can evaluate a wide range of relationships both linear and nonlinear efficiently and equitably even if nothing is known about the specific types of those relationships. And it compensates the disadvantages of Reshef's MIC that it can not be worked out accurately in polynomial time and is incapable of identifying the "local random" phenomenon.

## II. MEASURING CORRELATIONS

To evaluate bi-variable correlations, a proper topology structure is needed for partitioning samples. The most common used topology structures include histogram partition [11, 15] (Fig. 1-a) and Darbellay's recursive partition [12] (Fig. 1-b). Reshef can evaluate correlations theoretically based on histogram partition, but computational difficulty occurs when applied practically [15]. In this paper, $R_{A \times B}$ represents some kind of topology partition of rectangle region $A \times B$, and $R_{A \times B} = P_A \times P_B$ represents histogram partition. Darbellay can estimate mutual information of samples based on recursive partition, but correlations can not be measured correctly by MI. Combining the core thoughts of the two methods above, a topology partitioning method named "alternant recursive topology partition" is propsed by this paper, and a class of statistics for evaluating the properties of bi-variable correlations is put forward.

Alternant recursive topology partition (Fig. 1-c, represented by $R_\Omega^{r,c}$) is constructed similarly to Darbellay's recursive partition, where r, c denote the recursion depth of row and column respectively. First, histogram partition of scale of $1 \times 2$ (referred to as row-first-partition) or scale of $2 \times 1$ (referred to as column-first-partition) is applied to planar region $\Omega$, and then, column-first-partition $R_{A \times B}^{r-1,c}$ is applied to every rectangle $A \times B$ generated directly by row-first-partition of recursion depth of r, c, and row-first-partition $R_{A \times B}^{r,c-1}$ is applied to every rectangle $A \times B$ generated directly by column-first-partition of recursion depth of r, c. The above process is applied to each rectangle generated by sub-partition recursively, and $R_{A \times B}^{0,0}$ represents single rectang-le with no more sub-partition. For any rectangle $C \times D$ and recursion depth r, c, row-first-partition is denoted by $R_{C \times D}^{\bar{r},c}$ and column-first-partition by $R_{C \times D}^{r,\bar{c}}$, and let $R_{C \times D}^{0,c} = R_{C \times D}^{\bar{0},c}$, $R_{C \times D}^{r,0} = R_{C \times D}^{\bar{r},0}$. We call $R_{C \times D}^{r,c} \in R_{A \times B}^{r',c'}$ if $R_{C \times D}^{r,c}$ is a direct sub-partition of $R_{A \times B}^{r',c'}$, and any sub-partition belonging to $R_{C \times D}^{r,c}$ belongs to $R_{A \times B}^{r',c'}$.

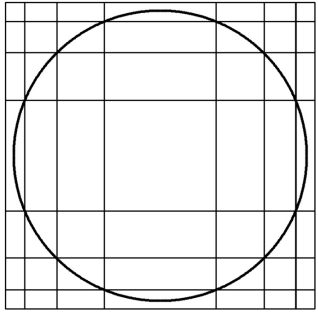 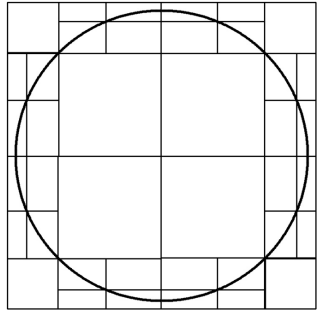 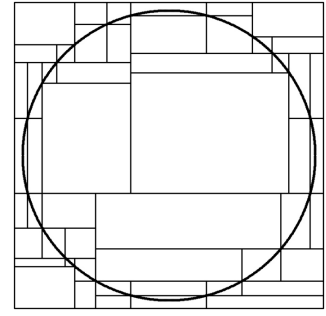

(a) Reshef's Histogram Partition     (b) Darbellay's Recursive Partition     (c) Alternant Recursive Topology Partition

Figure 1. Kinds of Topology Partitions of Standard Circle

All of the above topology partitioning methods can determine the topology structure uniquely, while, the same topology structure does not uniquely identify a partition. In order to meet the uniqueness of partition, Reshef requires that the mutual information of corresponding discrete distribution of histogram partition must be maximized, and Darbellay adopts the strategy of edge-even-partition when perform histogram partition of scale of $r \times r$ every recursive step. To make the alternant recursive topology partition also satisfy uniqueness, the "restricted divergence" [12] of $R_{\Omega}^{r,c}$ is required to be maximized, and such a partition is called "maximum restricted divergence alternant recursive topology partition", denoted by $\dot{R}_{\Omega}^{r,c}$. Fig. 1-a, 1-b, 1-c, respectively, represent Reshef's histogram partition of scale of 8×8 (thick line represents several adjacent partition lines), Darbellay's recursive partition of recursion depth of 3 and the maximum restricted divergence alternant recursive topology partition $\dot{R}_{\Omega}^{3,3}$ of standard circle. Beginning with definition 2.1, the computational method for $\dot{R}_{\Omega}^{r,c}$ and its maximum restricted divergence is derived by Theorem 2.1 and its corollary.

**Definition 2.1** Region Conditional Mutual Information is defined as follows:

$$I_{A \times B}(X,Y) = \iint_{A \times B} f(x,y \mid A \times B) \lg \frac{f(x,y \mid A \times B)}{f(x \mid A) f(y \mid B)} dxdy$$

where $A \times B$ represents any rectangle region, $f(x,y \mid A \times B)$ indicates joint conditional probability density function, $f(x \mid A)$ and $f(y \mid B)$ indicates edge conditional probability density function.

**Theorem 2.1** Let $R_{A \times B}$ represents an arbitrary partition of region $A \times B$, there are:

$$I_{A \times B}(X,Y) = \sum_{C \times D \in R_{A \times B}} \frac{P_{XY}(C \times D)}{P_{XY}(A \times B)} \lg \frac{P_{XY}(C \times D)}{P_X(C) P_Y(D)} - \lg \frac{P_{XY}(A \times B)}{P_X(A) P_Y(B)} + \sum_{C \times D \in R_{A \times B}} \frac{P_{XY}(C \times D)}{P_{XY}(A \times B)} I_{C \times D}(X,Y)$$

where $P_{XY}(\cdot \times \cdot)$ denotes joint regional probability, $P_X(\cdot)$ and $P_Y(\cdot)$ denote edge interval probability. The proof of Theorem 2.1 is shown as follows:

$$Proof: I_{A \times B}(X,Y) = \sum_{C \times D \in R_{A \times B}} P_{X,Y \mid A \times B}(C \times D) \iint_{C \times D} \frac{f(x,y \mid A \times B)}{P_{X,Y \mid A \times B}(C \times D)} \lg \frac{\frac{f(x,y \mid A \times B)}{P_{X,Y \mid A \times B}(C \times D)} P_{X,Y \mid A \times B}(C \times D)}{\frac{f(x)}{P_X(A)} \frac{f(y)}{P_Y(B)}} dxdy$$

$$= \sum_{C \times D \in R_{A \times B}} \frac{P_{XY}(C \times D)}{P_{XY}(A \times B)} \iint_{C \times D} f(x,y \mid C \times D) \lg \frac{f(x,y \mid C \times D) \frac{P_{XY}(C \times D)}{P_{XY}(A \times B)}}{\frac{f(x)}{P_X(C)} \frac{f(y)}{P_Y(D)} \frac{P_X(C) P_Y(D)}{P_X(A) P_Y(B)}} dxdy$$

$$= \sum_{C \times D \in R_{A \times B}} \frac{P_{XY}(C \times D)}{P_{XY}(A \times B)} \lg \frac{P_{XY}(C \times D)}{P_X(C) P_Y(D)} - \sum_{C \times D \in R_{A \times B}} \frac{P_{XY}(C \times D)}{P_{XY}(A \times B)} \lg \frac{P_{XY}(A \times B)}{P_X(A) P_Y(B)} + \sum_{C \times D \in R_{A \times B}} \frac{P_{XY}(C \times D)}{P_{XY}(A \times B)} I_{C \times D}(X,Y)$$

$$= \sum_{C \times D \in R_{A \times B}} \frac{P_{XY}(C \times D)}{P_{XY}(A \times B)} \lg \frac{P_{XY}(C \times D)}{P_X(C) P_Y(D)} - \lg \frac{P_{XY}(A \times B)}{P_X(A) P_Y(B)} + \sum_{C \times D \in R_{A \times B}} \frac{P_{XY}(C \times D)}{P_{XY}(A \times B)} I_{C \times D}(X,Y)$$

**Corollary 2.1** Let $R_\Omega$ represents an arbitrary partition of region $\Omega = R \times R$, there are:

$$I(X,Y) = \sum_{C \times D \in R_\Omega} P_{XY}(C \times D) \lg \frac{P_{XY}(C \times D)}{P_X(C)P_Y(D)} + \sum_{C \times D \in R_\Omega} P_{XY}(C \times D) I_{C \times D}(X,Y)$$

The proof of Corollary 2.1: It can be easily derived by mutual information $I(X,Y) = I_{\Omega = R \times R}(X,Y)$, slightly.

The right-hand side of the equation of Corollary 2.1 is a summation of two formulas. The former is the "restricted divergence", and the latter is called "residual divergence" by Darbellay, and he proved if $R_\Omega$ is detailed enough, then the mutual information asymptotically equals to the restricted divergence [12]. This is one of the reasons that we consider in this paper using alternant recursive topology partition to replace Reshef's histogram partition and restricted divergence to replace mutual information. Another important reason is about computational complexity that no polynomial time algorithm exists for computing Reshef's histogram partition, while, Theorem 2.1 provides us a polynomial time dynamic programming algorithm to work out the maximum restricted divergence alternant recursive topology partition.

Substituting $R_{A \times B} = R_{R \times R}^{r,c}$ into the Theorem 2.1, for $\forall R_{C \times D}^{0,0} \in R_{R \times R}^{r,c}$, let $I_{C \times D}(X,Y) = 0$, and then for $\forall R_{A \times B}^{p,q} \in R_{R \times R}^{r,c}$, let $I(R_{A \times B}^{p,q}) = I_{A \times B}(X,Y)$, easy to see that $I(R_{R \times R}^{r,c})$ is the restricted divergence of $R_{R \times R}^{r,c}(X,Y)$. On this basis, sample's maximum restricted divergence can be defined.

**Definition 2.2** For a finite sample set $D \subset R^2$ and positive integers r, c, row-first maximum restricted divergence $I^*(D, \bar{r}, c)$ is defined as follows:

$$I^*(D, \bar{r}, c) = I(\dot{R}_{R \times R}^{\bar{r},c}(X,Y)) = \max I(R_{R \times R}^{\bar{r},c}(X,Y))$$

where $(X,Y)$ is the discrete empirical distribution of D, $R_{R \times R}^{\bar{r},c}$ represents any row-first-partition of $(X,Y)$ with recursion depth $r$, $c$. Column-first maximum restricted divergence is defined similarly, denoted by $I^*(D, r, \bar{c})$.

The right-hand side of the equation of Theorem 2.1 is the optimal substructure of the left-hand side, therefore, recursive equation (1), (2) and (3) can be derived, and dynamic programming techniques can be used computing $I^*(D,r,c)$.

The time and space complexity of dynamic programming algorithm are the total number of sub-problems. Let $n = |D|$, it is easy to see that there are $O(n^4)$ different rectangular regions (rectangles containing the same points of D are considered the same one), and $r,c = O(\lg n)$ (noted below), thus the total number of sub-problems is $O(n^4 \lg^2 n)$, therefore, the time and space complexity are $O(n^4 \lg^2 n)$. In terms of computational complexity, it has been significantly improved when compared with Reshef's model in which no polynomial time algorithm exists. When sample size is large, this article rasterizes the sample evenly into discrete distribution of appropriate size firstly, then the calculation is executed, namely, the purpose of cutting down the computational cost can be achieved by reducing the precision appropriately.

$$I(R_{A \times B}^{0,0}) = 0, \quad \forall A \times B \in R^2 \tag{1}$$

$$\begin{aligned}
\text{MAX } I(R_{A \times B}^{\bar{r},c}) = \max_{P_A = \{A\}, P_B = \{B_1, B_2\}} \{ \\
\sum_{C \times D \in P_A \times P_B} \frac{P_{XY}(C \times D)}{P_{XY}(A \times B)} \lg \frac{P_{XY}(C \times D)}{P_X(C)P_Y(D)} + \\
\sum_{C \times D \in P_A \times P_B} \frac{P_{XY}(C \times D)}{P_{XY}(A \times B)} (\text{MAX } I(R_{C \times D}^{r-1,\bar{c}})) - \\
\lg \frac{P_{XY}(A \times B)}{P_X(A)P_Y(B)} \}, P_B \text{ is an arbitrary} \\
\text{bi-partition of } B
\end{aligned} \tag{2}$$

$$\begin{aligned}
\text{MAX } I(R_{A \times B}^{r,\bar{c}}) = \max_{P_A = \{A_1, A_2\}, P_B = \{B\}} \{ \\
\sum_{C \times D \in P_A \times P_B} \frac{P_{XY}(C \times D)}{P_{XY}(A \times B)} \lg \frac{P_{XY}(C \times D)}{P_X(C)P_Y(D)} + \\
\sum_{C \times D \in P_A \times P_B} \frac{P_{XY}(C \times D)}{P_{XY}(A \times B)} (\text{MAX } I(R_{C \times D}^{\bar{r},c-1})) - \\
\lg \frac{P_{XY}(A \times B)}{P_X(A)P_Y(B)} \}, P_A \text{ is an arbitrary} \\
\text{bi-partition of } A
\end{aligned} \tag{3}$$

**Definition 2.3** For positive integers r, c, row-first maximum restricted divergence upper limit $\dot{I}_{\bar{r},c}$ is defined as follows:

$$\dot{I}_{\bar{r},c} = \max_{\forall D} \{I^*(D, \bar{r}, c)\}$$

experiments show that $\dot{I}_{\bar{r},c} = I^*(L, \bar{r}, c)$, where L denotes any large-scale sample obeying linear distribution, therefore, the dynamic programming algorithm above can be used computing $\dot{I}_{\bar{r},c}$. Column-first maximum restricted divergence upper limit is defined similarly, denoted by $\dot{I}_{r,\bar{c}}$.

**Definition 2.4** For a finite sample set $D \subset R^2$, row-first characteristic matrix $M_r(D)$ is defined as an infinite matrix with entries as follows:

$$M_r(D)_{x,y} = I^*(D, \bar{x}, y) / \dot{I}_{\bar{x},y}$$

Column-first characteristic matrix is defined similarly, denoted by $M_c(D)$.

With the above basis, the definitions of statistics measuring the properties of bi-variable correlation will be given as follows.

**Definition 2.5** ARTMIC (Alternant Recursive Topology Maximum Information Coefficient)

$$\text{ARTMIC}(D) = \max_{x+y<=B(D)} \{ M_r(D)_{x,y}, M_c(D)_{x,y} \}$$

where $w(1) < B(D) \leq \lg_2 |D|^\varepsilon$, $0 < \varepsilon < 1$, in this paper we use $B(D) = \lg_2 |D|^{0.6}$, and $B(D)$ has the same meaning hereinafter. This statistic range from 0 to 1 and can be used to measure the strength of correlation.

**Definition 2.6** ARTMAS (Alternant Recursive Topology Maximum Asymmetry Score)

$$\text{ARTMAS}(D) = \max_{x+y<=B(D)} |M_r(D)_{x,y} - M_c(D)_{y,x}|$$

which ranges from 0 to 1 and no greater than ARTMIC. This statistic can be used to measure topological asymmetry of sample or deviation from monotonicity if there exists a functional relationship between two variables.

**Definition 2.7** ARTMEV (Alternant Recursive Topology Maximum Edge Value)

$$\text{ARTMEV}(D) = \max_{x+y<=B(D), x=1} \{ M_r(D)_{x,y}, M_c(D)_{y,x} \}$$

which ranges from 0 to 1 and no greater than ARTMIC. This statistic can be used to measure closeness to being a function.

**Definition 2.8** ARTMCN (Alternant Recursive Topology Minimum Cell Number)

$$\text{ARTMCN}(D) = \min_{x+y<=B(D)} \{ x+y-2:$$

$$M_r(D)_{x,y} \geq (1-\varepsilon)\text{ARTMIC}(D)$$

$$\text{or } M_c(D)_{x,y} \geq (1-\varepsilon)\text{ARTMIC}(D)$$

$$\}, \quad 0 < \varepsilon < 1$$

the value of which is non-negative integers. This statistic can be used to measure complexity of relationship, the greater the value the more complex relationship.

**Definition 2.9** ARTLRD (Alternant Recursive Topology Local Random Degree)

$$\text{ARTLRD}(D) = \max_{x+y<=B(D)} \{ M_r(D)_{x,y}, M_c(D)_{x,y} \} - \max_{\substack{x+y<=B(D) \\ x,y>=B(D)*q}} \{ M_r(D)_{x,y}, M_c(D)_{x,y} \}$$

where q ranges from 0 to 0.5, in this paper we use q=0.3. This statistic ranges from 0 to 1 and no greater than ARTMIC, and can be used to measure "Local Random Degree" of relationship.

The five statistics given by definition 2.5 to 2.9 are referred to as ART statistics in this paper. ART statistics can be used discovering correlations as follows: First, for each pair of variables, compute their ART statistics; then filter out specific pairs of variables by using ART statistics as conditions, these pairs are the expected relationships.

## III. PROPERTIES OF ART AND EXPERIMENTAL VERIFICATION

Definition 2.5 to 2.9 give some properties of ART, and this section further discusses other properties of ARTMIC and ARTLRD, and verifies them by experiments.

As sample size increasing, ARTMIC asymptotically satisfies the following properties: (a) ARTMIC of any noiseless nowhere-constant functional relationship converges to 1; (b) ARTMIC of statistically independent samples converges to 0; (c) For any type of relationship, the value of ARTMIC is inversely proportional to noise level; (d) For different types of nowhere-constant functional relationships with same noise level, their ARTMIC values are about the same.

TABLE I lists the results of ART and Reshef's statistics of some typical bi-variable relationships. The size of the sample points drawn from those relationships analyzed in this experiment is 1080. In TABLE I, $R_1$ represents a statistically independent relationship, $R_2$ represents a nowhere-flat, nowhere-vertical linear relationship, $R_3$ represents a sinusoidal relationship whose definition domain is $[0, 5\pi]$, $R_4$ represents a parabolic relationship, $R_5$ represents a standard circle relationship. Comparative analysis of TABLE I shows that the two classes of statistics are largely consistent except that ARTMCN is smaller than MCN, the reason of this exception is that ARTMCN ranges from 0, while, the MCN ranges from 2, and it is considered more reasonable in this paper that the complexity of the simplest relationship equals to 0. Both ARTMIC and MIC satisfy the properties (a-c), but for property d, ARTMIC performs better than MIC knew by comparing the results of $R_2$, $R_3$, $R_4$ in TABLE I. Especially when perturbation is large, the value of ARTMIC for functional relationships of different types is more stable, showing that ARTMIC can treat relationships of different types more equitably.

ARTLRD is a completely new statistic proposed in this paper for measuring the degree of local random phenomenon in which strong correlation as a whole and large random in the local occur at the same time (shown by Fig. 2-c to 2-f), the more obvious this phenomenon, the greater the value of ARTLRD. The original intention of proposing ARTLRD is to solve the erroneous identification problem of MIC and ARTMIC. As shown in TABLE II (sample a to f, respectively, illustrated by Fig. 2-a to 2-f, the size of each sample is 1080), when local random phenomenon is apparent (sample c to f), MIC and ARTMIC will ignore this phenomenon and identify them as strong correlations mistakenly, such as values of MIC and ARTMIC of sample c to f are all equal to 1, this is contrary to common sense. ARTLRD can solve this problem, real strong correlation exists when ARTMIC is large and ARTLRD is small (sample a); strong local random phenomenon exists when both ARTMIC and ARTLRD are large, and the value of ARTLRD reflects the degree of local random phenomenon (sample c-f); ARTLRD must be small and sample can be considered statistically independent when ARTMIC is small (sample b); it is known by definition that ARTLRD can not be large when ARTMIC is small. Therefore, ARTLRD can effectively evaluate local random phenomenon and assist in solving the erroneous identification problem of MIC and ARTMIC.

TABLE I. ART AND RESHEF'S STATISTICS OF TYPICAL RELATIONSHIPS

| Relationship | Perturbation | Correlation Strength | | | Topological Asymmetry | | Closeness to Being a Function | | Complexity | | Local Random |
|---|---|---|---|---|---|---|---|---|---|---|---|
| | | ARTMIC | MIC | Pearson's r | ARTMAS | MAS | ARTMEV | MEV | ARTMCN | MCN | ARTLRD |
| $R_1$ | - | 0.107 | 0.133 | -0.020 | 0.021 | 0.008 | 0.107 | 0.133 | 4.000 | 5.910 | 0.016 |
| $R_2$ | 0.0 | 1.000 | 1.000 | 1.000 | 0.000 | 0.000 | 1.000 | 1.000 | 0.000 | 2.000 | 0.000 |
| | 0.1 | 0.915 | 0.947 | 0.981 | 0.008 | 0.021 | 0.915 | 0.947 | 1.000 | 2.585 | 0.134 |
| | 0.3 | 0.699 | 0.685 | 0.865 | 0.028 | 0.036 | 0.699 | 0.685 | 1.000 | 4.585 | 0.239 |
| $R_3$ | 0.0 | 1.000 | 1.000 | 0.000 | 0.638 | 0.843 | 1.000 | 1.000 | 2.000 | 3.302 | 0.139 |
| | 0.1 | 0.894 | 0.965 | -0.003 | 0.620 | 0.798 | 0.894 | 0.965 | 2.000 | 4.169 | 0.199 |
| | 0.3 | 0.757 | 0.776 | 0.032 | 0.501 | 0.636 | 0.757 | 0.776 | 3.000 | 4.700 | 0.307 |
| $R_4$ | 0.0 | 1.000 | 1.000 | 0.000 | 0.539 | 0.689 | 1.000 | 1.000 | 1.000 | 2.585 | 0.066 |
| | 0.1 | 0.900 | 0.954 | -0.006 | 0.469 | 0.654 | 0.900 | 0.954 | 1.000 | 3.322 | 0.193 |
| | 0.3 | 0.700 | 0.627 | -0.010 | 0.346 | 0.396 | 0.700 | 0.627 | 2.000 | 3.807 | 0.290 |
| $R_5$ | 0.0 | 0.625 | 0.712 | 0.000 | 0.000 | 0.000 | 0.461 | 0.322 | 3.000 | 4.640 | 0.000 |
| | 0.1 | 0.479 | 0.527 | 0.002 | 0.005 | 0.018 | 0.414 | 0.280 | 3.000 | 4.390 | 0.000 |
| | 0.3 | 0.293 | 0.314 | 0.004 | 0.005 | 0.014 | 0.289 | 0.153 | 2.000 | 5.170 | 0.000 |

TABLE II. MIC, ARTMIC, ARTLRD OF KINDS OF SAMPLES

| | a | b | c | d | e | f |
|---|---|---|---|---|---|---|
| MIC | 1.000 | 0.133 | 1.000 | 1.000 | 1.000 | 1.000 |
| ARTMIC | 1.000 | 0.107 | 1.000 | 1.000 | 1.000 | 1.000 |
| ARTLRD | 0.000 | 0.006 | 0.739 | 0.558 | 0.480 | 0.382 |

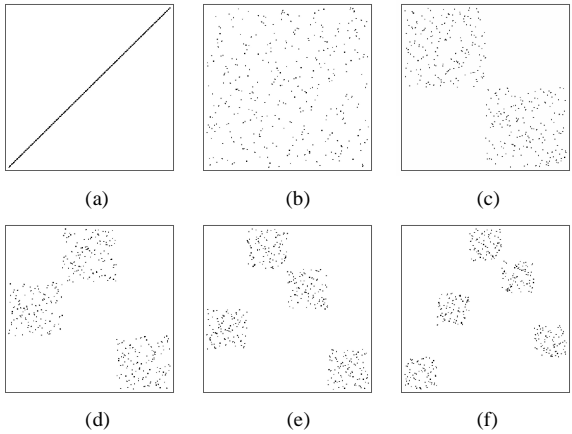

(a) (b) (c)
(d) (e) (f)

Figure 2. Samples with Local Random Phenomenon of Different Levels

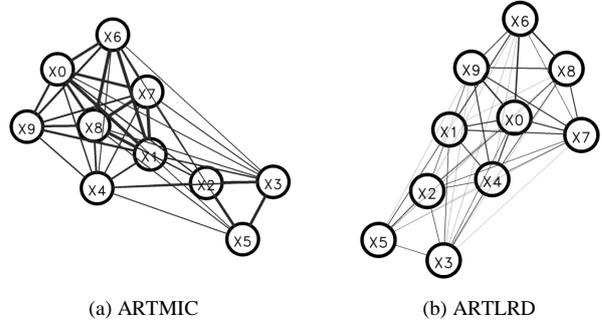

(a) ARTMIC  (b) ARTLRD

Figure 3. ART Statistics between Each Pair of Variables

## IV. APPLICATION OF ART

In this paper, a dataset of 10 American classical indexes with data collected from 1959-1 to 2013-5 is analyzed by ART. This dataset contains 641 sample points, which can be downloaded at www.shujuhui.com. These 10 indexes ($X_0$ – $X_9$) are: Employment-population Ratio, Total Employed Population, Purchasing Managers' Index, Output Index, Price Index, New Orders Index, Individual Reserve Deposits per Disposable Income, Sum of Individual Deposits, Total Rental Income Adjusted by Capital Loss, Consumer Expenditures Price Index.

Fig. 3-a and Fig. 3-b represent the ARTMIC and ARTLRD between each pair of variables ($X_0$ – $X_9$) respectively, the circles in the figures denote variables and the thickness of each line connecting two circles indicates the value of specific ART statistic between correspondent variables, and if two circles are not connected, then the ART statistic between correspondent variables is 0 or very low, similarly, figures illustrating other ART statistics can be drawn. ART can be used mining novel correlations, for example, line connecting $X_1$ and $X_6$ in Fig. 3-a is thick (correspondent ARTMIC equals to 0.87), line connecting $X_1$ and $X_6$ in Fig. 3-b is of medium width (correspondent ARTLRD equals to 0.24), in addition, other ART statistics between $X_1$ and $X_6$ are : 0.049(ARTMAS), 0.87(ARTMEV), 1.00(ARTMCN), therefore, it can be inferred that strong functional relationship exists between $X_1$ and $X_6$ by the large values of ARTMIC and ARTMEV, and this relationship has a certain topological symmetry by the small value of ARTMAS, and the relationship is not very complex by ARTMCN equaling to 1, and this relationship has a certain degree of local random phenomenon by ARTLRD equaling to 0.24. Only depending on data, ART provides lots of prop-

erties of the relationship between $X_1$ and $X_6$, which can assist scientists to make certain assumptions or conjectures and provide a basis for further studying the relationship. Actually, the sample image of $X_1$ and $X_6$ (Fig. 4-a, processed by a topological transformation, horizontal and vertical coordinates of sample points are $X_1$ and $X_6$ respectively) shows that $X_1$ and $X_6$ these two seemingly unrelated indexes obey sine function relationship approximately. Lots of bi-variable correlations have been mined by ART just like the way above, and the correlations shown in Fig. 4 are only small part of them.

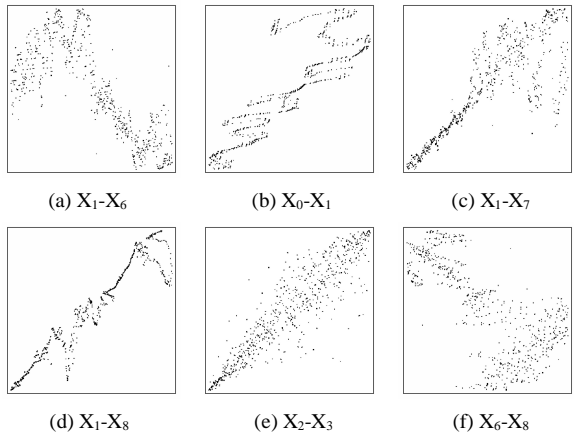

(a) $X_1$-$X_6$  (b) $X_0$-$X_1$  (c) $X_1$-$X_7$

(d) $X_1$-$X_8$  (e) $X_2$-$X_3$  (f) $X_6$-$X_8$

Figure 4. Sample Images of Bi-variable Correlations

## V. CONCLUSIONS

A class of statistics named ART is proposed in this paper to evaluate the properties of correlation between two variables. ART can evaluate a wide range of relationships both linear and nonlinear efficiently. And it compensates the disadvantages of Reshef's model that it can not be worked out accurately in polynomial time and is incapable of identifying the local random phenomenon.

Just like MIC, the anti-noise ability of ARTMIC is not strong when sample size is small, in addition, no method exists so far for measuring and discoverring multivariable correlations, and these will be the focus of future research.

ACKNOWLEDGMENT

This paper is supported by the National Natural Science Foundation of China under Grant No. 61273314, the Special Science and Technology Foundation for Changsha (No. k1205014-11) and the Fundamental Research Funds for the Central Universities of Central South University (No. 2012zzts095).

REFERENCES

[1] Science Staff, "Challenges and opportunities," Science, vol. 331, no. 6018, pp. 692-693, 2011.

[2] B. Hanson, A. Sugden, and B Alberts, "Making data maximally available," Science, vol. 331, no. 6018, pp. 649, 2011.

[3] J. Carpenter, "May the best analyst win," Science, vol. 331, no. 6018, pp. 698-699, 2011.

[4] J. Overpeck, G. Meehl, S. Bony, and D. Easterling, "Climate data challenges in the 21st century," Science, vol. 331, no. 6018, pp. 700-702, 2011.

[5] O. Reichman, M. Jones, and M. Schildhauer, "Challenges and opportunities of open data in ecology," Science, vol. 331, no. 6018, pp. 703-705, 2011.

[6] H. Akil, M. Martone, and D. Van Essen, "Challenges and opportunities in mining neuroscience data," Science, vol. 331, no. 6018, pp. 708-711, 2011.

[7] J. Rodgers, and W. Nicewander, "Thirteen ways to look at the correlation coefficient," The American Statistician, vol. 42, no. 1, pp. 59-66, 1988.

[8] T. Speed, "A correlation for the 21st century," Science, vol. 334, no. 6062, pp. 1502–1503, 2011.

[9] T. Cover, and J. Thomas, Elements of information theory, 2nd ed., John Wiley & Sons Inc: New Jersey, 2006.

[10] E. Linfoot, "An information measure of correlation," Information and Control, vol. 1, no. 1, pp. 85-89, 1957.

[11] R. Steuer, J. Kurths, C. Daub, J. Weise, and J. Selbig, "The mutual information: detecting and evaluating dependencies between variables," Bioinformatics, vol. 18, no. 2, pp. 231–240, 2002.

[12] G. Darbellay, and I. Vajda, "Estimation of the information by an adaptive partitioning of the observation space," IEEE Transactions on Information Theory, vol. 45, no. 4, pp. 1315–1321, 1999.

[13] A. Kraskov, H Stogbauer, and P Grassberger, "Estimating mutual information," Physical Review, vol. 69, no. 6, pp. 066138, 2004.

[14] Y. Moon, B. Rajagopalan, and U. Lall, "Estimation of mutual information using kernel density estimators," Physical Review, vol. 52, no. 3, pp. 2318–2321, 1995.

[15] D. Reshef, Y. Reshef, H. Finucane, S. Grossman, G. Mcvean, P. Turnbaugh, E. Lander, M. Mitzenmacher, and P. Sabeti, "Detecting novel associations in large data sets," Science, vol. 334, no. 6062, pp. 1518–1524, 2011.

[16] N. Simon, and R. Tibshirani, 2012, "Comment on "Detecting novel associations in large data sets" by Reshef et. al (Science DEC 2011),". [Online]. Available: http://www-stat.stanford.edu/~tibs/reshef/comment.pdf